\documentclass[a4paper,11pt]{article}
\usepackage{jinstpub} 
\usepackage{lineno}

\proceeding{Topical Workshop on Electronics for Particle Physics - TWEPP 2026\\
September 28, 2026 to October 2, 2026\\
Castelldefels, Barcelona, Spain}

\title{A Scalable Two-Stage Back-End Architecture for 400GbE Data Acquisition Systems}

\author[1]{A. Perro,\note{Corresponding author.}}
\author{P. Durante,}
\author{T. Colombo}
\affiliation{CERN\\Geneva, Switzerland}
\emailAdd{alberto.perro@cern.ch}

\abstract{A primary challenge in High Energy Physics (HEP) data acquisition is the aggregation of numerous low-speed (10~Gb/s) radiation-hard data links into high-speed back-end interfaces. While high-tier FPGAs traditionally perform this task, transceiver density and speed in modern devices can limit cost-effective scaling. This study proposes a two-stage architecture that utilizes mid-range FPGAs for decoding and Ethernet conversion, followed by a Microchip META-DX2+ Ethernet PHY that aggregates these streams into a 400~GbE interface. This study evaluates the design’s performance, validating it as a scalable and cost-effective architecture for next-generation HEP data acquisition systems.}

\notoc

\begin{document}
\maketitle
\flushbottom
\section{Introduction}
\label{sec:intro}
\paragraph{Readout Architectures at LHC}
Data Acquisition (DAQ) systems across the Large Hadron Collider (LHC) experiments share many architectural principles.
Detectors are large machines composed of multiple sub-detectors that produce aggregate data rates on the order of tens of terabits per second (Tb/s).
These machines are placed in experimental areas which are exposed to ionizing radiation, therefore they require purposely designed electronics.
Moreover, the detectors are physically large, spread across tens of meters in all directions and data links must collect data from detector areas with highly uneven occupancy.

As an example, the LHCb experiment~\cite{pisani_design_2023} currently utilizes approximately 11,000~Versatile Link (VL) links~\cite{baron_gbt_2009} operating at 4.8~Gb/s.
Most of these links use CERN's custom radiation-hard optical protocol called GBT.
The links terminate at multiple custom high-tier FPGA readout cards which aggregate, process, and transfer the incoming data via PCI Express (PCIe)~\cite{cachemiche_pciebased_2016}\cite{durante_100_2015} to the Event Builder farm nodes.
A total of 520 cards is used in LHCb, 445 of which are dedicated to readout.

The readout cards are hosted in groups of three within Event Builder servers.
Each chassis can accommodate up to eight dual-slot PCIe cards, designed specifically to support GPU-accelerated computing.
Because event fragments are distributed across the entire server farm, they must be gathered and assembled prior to real-time reconstruction and trigger selection.
The servers are interconnected via a high-speed InfiniBand HDR (200~Gb/s) network designed for a 32~Tb/s aggregate bandwidth.
The system relies on Remote Direct Memory Access (RDMA) and specialized event-building algorithms~\cite{pisani_design_2023} to sustain the required throughput.

The key features and constraints of the described architecture are:
\begin{enumerate}
    \item Many links with custom radiation-had protocols, because of routing and environmental constraints.
    \item High order of data aggregation to optimize cost per unit throughput.
    \item High-bandwidth inter-server networking for event building, prior to software filtering.
\end{enumerate}

For the High-Luminosity LHC (HL-LHC) era, the LHCb sub-detectors and Front-End Electronics (FEE) will be upgraded to handle significantly higher instantaneous luminosities.
The detector links will transition to the Low Power GigaBit Transceiver (lpGBT) technology~\cite{moreira_lpgbt_2025} reaching up to 10.24~Gbps per link.
For Run~5 (2036--2041), LHCb foresees approximately 30,000 readout links and an aggregate throughput exceeding 300~Tb/s~\cite{colombo_online_2024}, necessitating a novel DAQ architecture aligned with modern computing paradigms.

\paragraph{Industry Trends}
Modern DAQ systems rely on Commercial Off-The-Shelf (COTS) components, such as enterprise compute servers and network switches. this approach can lower infrastructure cost through economies of scale, but it binds DAQ architectures to broader industry hardware trends.

In inter-server communications, Ethernet has established itself as the dominant standard. Modern Ethernet Network Interface Cards (NICs) achieve bandwidths of up to 800~Gb/s, together with switching ASICs scaling to 102.4~Tb/s.
Protocols such as RDMA over Converged Ethernet (RoCEv2) and the emerging Ultra Ethernet standard provide InfiniBand-class latency and transport performance within a more open multi-vendor ecosystem.

Intra-server interconnects are dominated by PCIe.
However, the compute requirements driven by artificial intelligence workloads have pushed power delivery and thermal envelopes well beyond the capabilities of standard dual-slot, Full-Height Full-Length (FHFL) PCIe add-in cards.
To dissipate kilowatts of heat within dense server chassis, major vendors have shifted toward dedicated mezzanine module architectures, such as SXM and OAM.
For these reasons, commercial enterprise servers capable of accommodating multiple standard FHFL PCIe cards are becoming increasingly scarce and expensive to procure.

To address these challenges, this paper presents a cost-optimized DAQ architecture built on high-bandwidth Ethernet. By relocating the FPGA readout boards into dedicated standalone enclosures, the design completely decouples the readout layer from enterprise form-factor constraints.

\section{Proposed Solution}
\label{sec:design}
\paragraph{Media and Protocol Conversion}
A primary requirement of any DAQ link aggregator is the media and protocol conversion.
FPGAs are required to decode the custom radiation-hard protocols and aggregate them in faster Ethernet links.
Modern high-tier FPGAs provide transceivers capable of reaching up to 112~Gb/s per lane.
These devices optimize for per-lane throughput at the expense of total transceiver count and configurability.
Detector readout links operate at 10~Gb/s and connecting them directly to high-speed SerDes leads to high transceiver under-utilization and they might operate outside the transceivers' supported line-rate range.

\begin{figure}[htbp]
    \centering
    \begin{minipage}[c]{0.48\textwidth}   
        \centering
        \includegraphics[width=\linewidth]{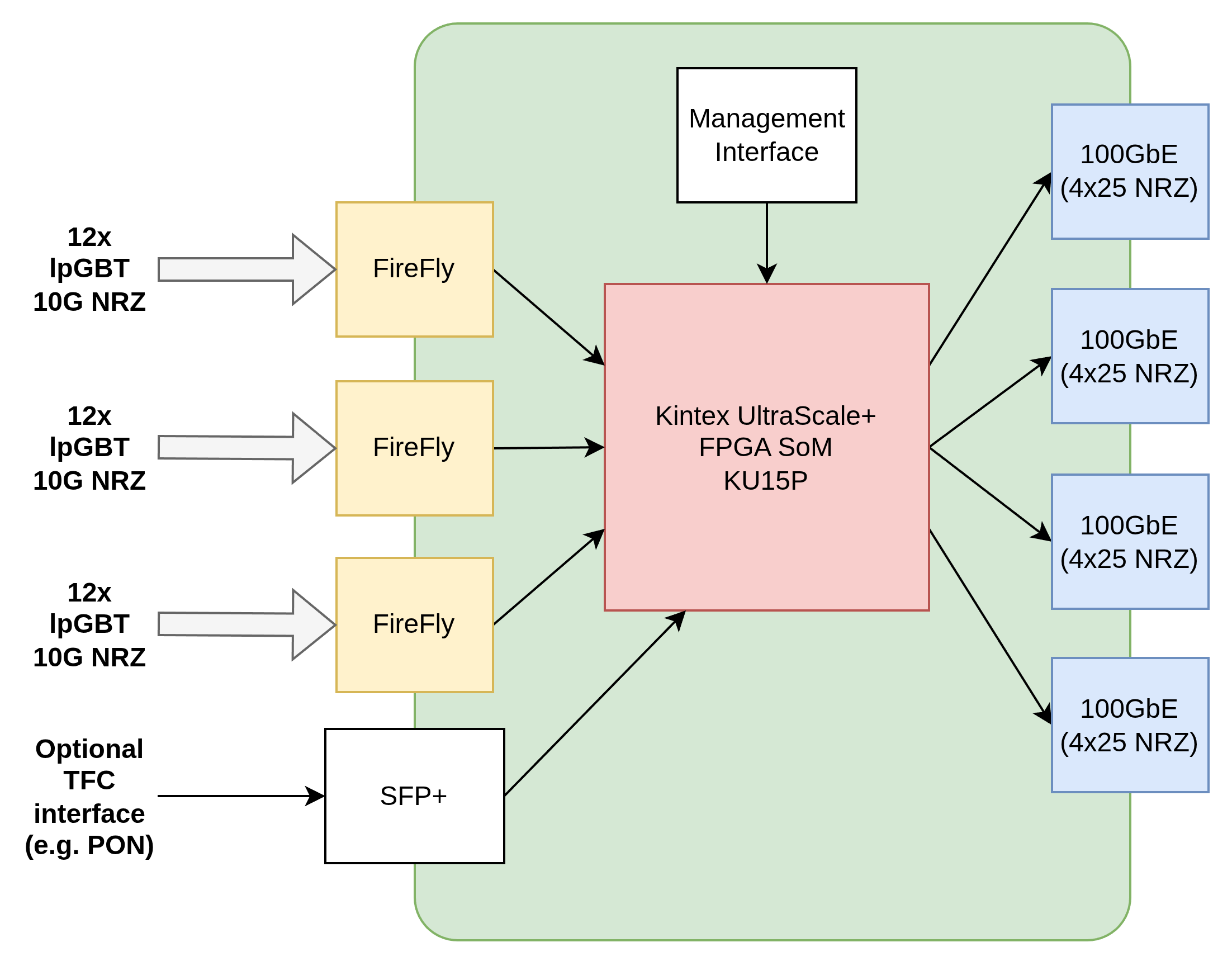}
        \caption{First-stage media and protocol conversion based on an AMD/Xilinx Kintex UltraScale+ FPGA. The device aggregates 36 lpGBT links (10.24~Gb/s) into four 100~GbE interfaces.}
        \label{fig:fpga}
    \end{minipage}%
    \hfill
    \begin{minipage}[c]{0.4\textwidth}
        \centering
        \includegraphics[width=\linewidth]{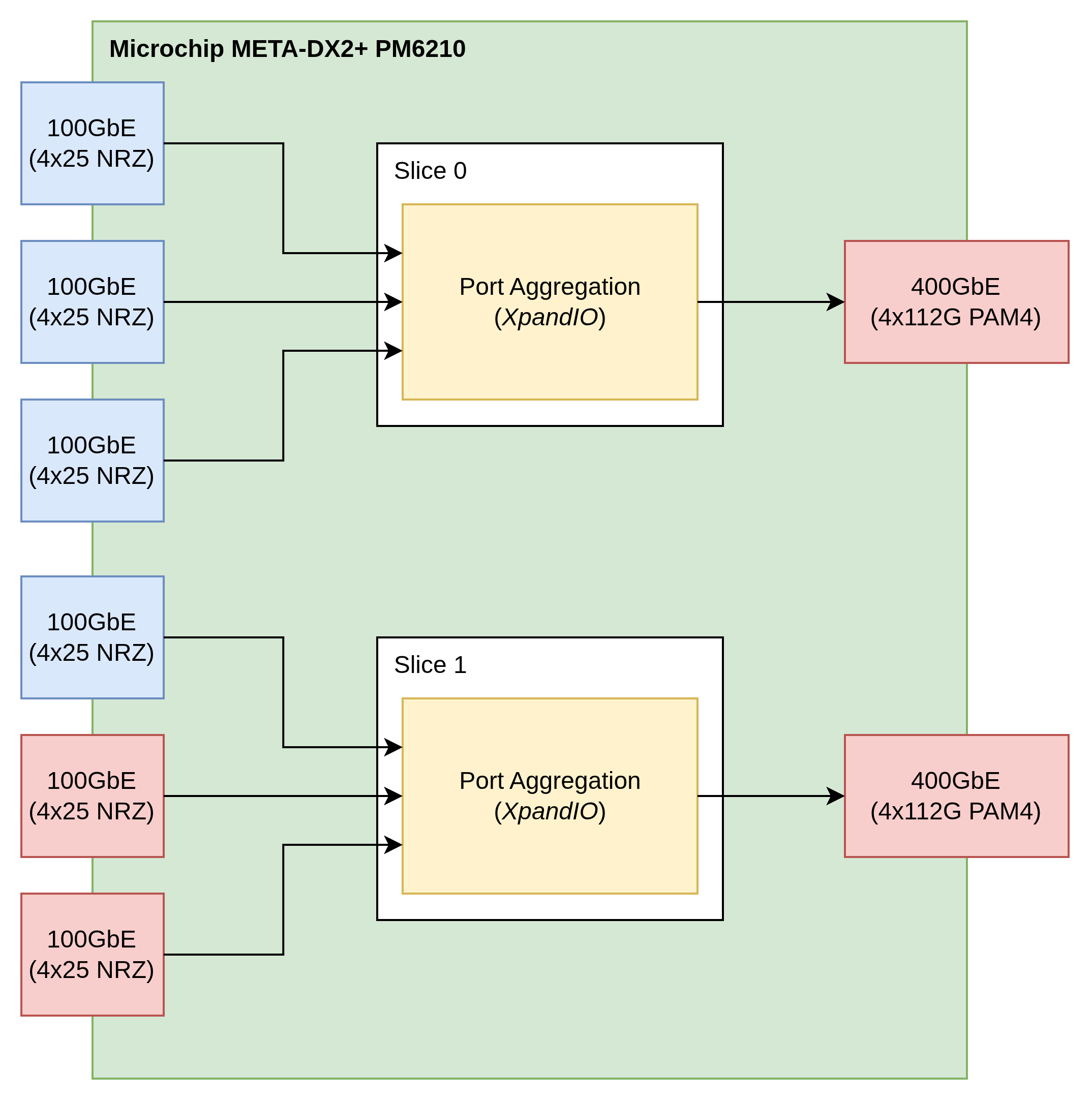}
        \caption{Port aggregation architecture in the META-DX2+ PM6210 PHY. Line SerDes are shown in blue and host SerDes in red. SerDes allocation can be dynamically routed using the internal \textit{ShiftIO} crossbar.}
        \label{fig:metaconf}
    \end{minipage}%
\end{figure}

In contrast, mid-tier FPGAs feature previous-generation transceivers (up to 28~Gb/s NRZ), which provide a balanced trade-off between transceiver density, package cost, and compatibility.
As illustrated in Figure~\ref{fig:fpga}, an AMD/Xilinx Kintex UltraScale+ FPGA can ingest 36 lpGBT links (10.24~Gb/s) and aggregate them into four 100~GbE interfaces. A proof-of-concept demonstrating this conversion stage was previously detailed in~\cite{perro_lowcost_2025}.

The mid-tier FPGA implements 100~GbE using the 100GBASE-KR4 standard, which utilizes four 25.78~Gb/s NRZ lanes.
Modern datacenter switches operating at 400~GbE and 800~GbE natively rely on 106~Gb/s PAM4 SerDes.
Interfacing switch ports directly with legacy 25~Gb/s lanes leads to the same under-utilization discussed above and increases total infrastructure cost.
To bridge this rate mismatch, we introduce a secondary aggregation stage based on a dedicated smart Ethernet PHY.

\paragraph{Ethernet Aggregation}
For the secondary aggregation stage, we chose the Microchip META-DX2+ PM6210 PHY~\cite{_metadx_nodate}. This device integrates 32 SerDes lanes operating at up to 106~Gb/s PAM4 and provides port aggregation capabilities, commercially referred to as \textit{XpandIO}.

The PHY is partitioned into two independent, symmetrical slices.
Each slice can aggregate up to eight incoming Ethernet streams into fewer, higher-rate channels.
In our implementation, illustrated in Figure~\ref{fig:metaconf}, each slice aggregates three 100GBASE-KR4 streams into a single 400GBASE-KR4 output interface.

The \textit{XpandIO} aggregation mechanism operates at OSI Layer~2~\cite{zimmermann_osi_1980}. Incoming frames are parsed and encapsulated with an IEEE 802.1Q VLAN tag corresponding to the ingress physical port before being multiplexed into the 400~GbE stream.
Downstream receivers must support VLAN de-encapsulation to separate the data streams.
While frame processing and tagging introduce minor packet overhead, back-pressure is avoided by design: in this configuration, the aggregate source throughput utilizes at most 75\% (300~Gb/s) of the 400~Gb/s uplink bandwidth, leaving ample headroom for framing overhead and line-rate burst tolerance.

The following section describes the experimental testbench used to evaluate this dual-stage aggregation architecture and discusses the performance results.

\section{Benchmarks}
\label{sec:tb}
\paragraph{Setup}
The testbench was configured to evaluate the aggregation of three 100~GbE streams into a single 400~GbE stream.
The source server features an AMD EPYC 9334 CPU, 768~GB DDR5 RAM (4800~MT/s), and $3\times$~NVIDIA ConnectX-5 100~GbE NICs.
The destination server is equipped with an Intel Xeon Gold 6454S CPU, 512~GB DDR5 RAM (4800~MT/s), and $1\times$~NVIDIA ConnectX-8 $2\times 400$~GbE NIC.
Both nodes run RHEL~9.5 (Linux kernel~5.14) with DPDK v22.11.2510.2.1, NIC firmware v40.47.1088, and NVIDIA OFED v25.10-1.7.1.

The NICs connect to a META-DX2+ Development Kit via Direct Attach Copper (DAC) cables.
The kit contains two onboard META-DX2+ PHYs interconnected via eight SerDes lanes.
The second PHY operates in Analog Retimer mode, providing a direct passthrough that exposes these eight SerDes lanes to the OSFP cages.

\paragraph{Methodology}
Synthetic traffic generators simulate the incoming detector data streams. Three independent, parallel UDP packet generators transmit 8~kB frames at a target rate of 100~Gb/s per stream. Each packet payload embeds a unique Source~ID and a 64-bit monotonically increasing sequence counter to track packet order and losses.

\begin{figure}[htbp]
    \centering
    \includegraphics[width=0.45\linewidth]{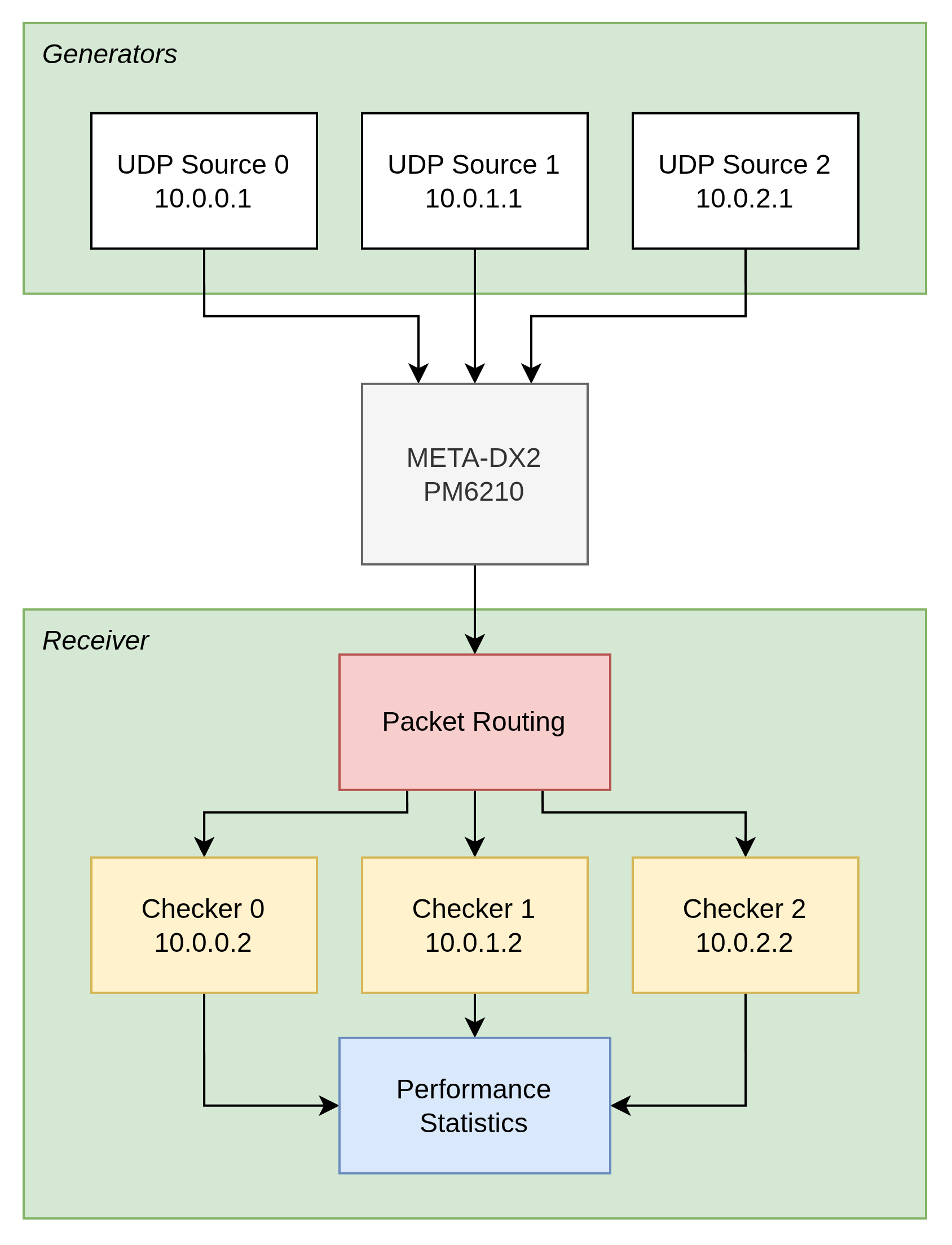}
    \caption{Architecture of the benchmarking software. On the source server, three independent DPDK generators (white) produce 100~Gb/s streams. On the destination server, an ingress dispatcher thread (red) demultiplexes packets by Source~ID to dedicated checker threads (yellow) for integrity checks. Global counters and throughput metrics are gathered by a central collector thread (blue).}
    \label{fig:tbsw}
\end{figure}

The receiver application, shown in Figure~\ref{fig:tbsw}, consists of:
\begin{itemize}
    \item A \textit{routing thread} (shown in red) that inspects the ingress packets and dispatches each stream to a dedicated downstream thread based on its Source~ID.
    \item Three independent \textit{checker threads} (shown in yellow), each allocated to a specific source stream, responsible for verifying ordering, packet loss, and calculating stream statistics.
    \item A \textit{collector thread} (shown in blue) that aggregates application-wide performance counters.
\end{itemize}

Both generators and receiver are developed using the Data Plane Development Kit (DPDK) framework~\cite{_dpdk_nodate} to bypass kernel networking overhead and achieve sustained line-rate throughput.

\paragraph{Results}
The results, summarized in Table~\ref{tab:res}, demonstrate that a single slice of the META-DX2+ PHY aggregates three concurrent 100~GbE streams without packet drops, reordering, or observable back-pressure.
The metrics are continuously captured over 5~TB of data and 610 million packets, resultin in a upper bound on the bit error rate of \(\text{BER} < 2 \times 10^{-13}\).

\begin{table}[htbp]
\centering
\caption{Throughput and packet loss measurements for a single META-DX2+ slice across 5~TB of payload data (610 million packets). All streams achieved expected line rates without packet loss or back-pressure.}
\label{tab:res}
\vspace{0.2cm}
\begin{tabular}{lrr}
\hline
\textbf{Stream} & \textbf{Throughput (Gb/s)} & \textbf{Packet Loss (\%)} \\
\hline
Source 0        & 98.8                       & 0.0                       \\
Source 1        & 99.1                       & 0.0                       \\
Source 2        & 98.9                       & 0.0                       \\
\hline
\textbf{Total}  & \textbf{296.8}             & \textbf{0.0}              \\
\hline
\end{tabular}
\end{table}

These measurements confirm that the PHY meets the performance and throughput requirements.
The second PHY slice was separately validated at the signal-integrity level using specialized high-frequency RF test fixtures supplied by Microchip.
Due to the physical layout and routing constraints of the evaluation kit, throughput benchmarks could not be executed simultaneously on both slices.

\section{Architecture Cost Evaluation}
This two-stage architecture was specifically designed to optimize the cost of the readout infrastructure.
Table~\ref{tab:cost} compares the unit cost per lpGBT link across a traditional PCIe-based baseline, a COTS switch-based network design, and the proposed META-DX2+ architecture using an AMD Kintex UltraScale+ KU15P FPGA.

Cost per LpGBT link ingested was chosen as a metric since it can be scaled to different layouts without being dependent on the specific hardware configuration.
The FPGA chosen for the Ethernet Readout is the Kintex Ultrascale Plus KU15P.
PCB and extras costs are estimated from similar designs.

Replacing intermediate COTS switches with the dedicated META-DX2+ PHY reduces network cost by nearly 38\%, with an overall cost reduction of 28.5\% per link compared to the PCIe-based solution.

\begin{table}[h]
\label{tab:cost}
\caption{Cost comparison per ingested lpGBT link across readout architectures.}
\vspace{0.2cm}
\centering
\begin{tabular}{|l|r|r|r|}
\hline
                          & \multicolumn{1}{c|}{\textbf{\begin{tabular}[c]{@{}c@{}}PCIe-based\\ Readout\end{tabular}}} & \multicolumn{1}{c|}{\textbf{\begin{tabular}[c]{@{}c@{}}COTS Switch\\ Network Readout\end{tabular}}} & \multicolumn{1}{c|}{\textbf{\begin{tabular}[c]{@{}c@{}}META-DX2+\\ Network Readout\end{tabular}}} \\ \hline
FPGA                      & 183.33                                                                                     & 75.86                                                                                               & 75.86                                                                                             \\ \hline
Network                   & 17.49                                                                                      & 91.26                                                                                               & 56.63                                                                                             \\ \hline
PCB and extras            &                                                                                            & 11.11                                                                                               & 11.11                                                                                             \\ \hline
\textbf{TOTAL (\$/lpGBT)} & \textbf{200.82}                                                                            & \textbf{178.23}                                                                                     & \textbf{143.6}                                                                                    \\ \hline
Savings                   &                                                                                            & 11.25\%                                                                                             & 28.49\%                                                                                           \\ \hline
\end{tabular}
\end{table}

\section{Conclusions}
This work demonstrates that a two-stage aggregation readout built on COTS Ethernet is a viable approach.
The META-DX2+ Ethernet PHY ensures a future-proof interface by supporting the latest high-speed Ethernet standards.
Overall, the design offers a cost-effective and competitive alternative to solutions based on high-tier FPGAs.
Furthermore, this study shows that standard UDP streams can sustain high-throughput readout systems without the need for RDMA.

\acknowledgments

The authors would like to acknowledge the support from the CERN LHCb Online Team, CERN EP R\&D WP 9.3, and ECFA DRD7.5b collaborations.

\bibliographystyle{JHEP.bst}
\bibliography{biblio.bib}

\end{document}